\documentclass[onecolumn]{aastex63}

\usepackage[english]{babel}
\usepackage{amsmath,amsfonts,amsthm}
\usepackage{amssymb}
\usepackage{graphicx}

\begin{document}
\title{The Diversity of Environments Around Luminous Quasars at Redshift $z\sim6$}

\author{Keven Ren}
\affiliation{School of Physics, The University of Melbourne, Parkville, Victoria, Australia}
\affiliation{ARC Centre of Excellence for All Sky Astrophysics in 3 Dimensions (ASTRO 3D)}

\author{Michele Trenti}
\affiliation{School of Physics, The University of Melbourne, Parkville, Victoria, Australia}
\affiliation{ARC Centre of Excellence for All Sky Astrophysics in 3 Dimensions (ASTRO 3D)}

\author{Madeline A. Marshall}
\affiliation{School of Physics, The University of Melbourne, Parkville, Victoria, Australia}
\affiliation{ARC Centre of Excellence for All Sky Astrophysics in 3 Dimensions (ASTRO 3D)}
\affiliation{National Research Council of Canada, Herzberg Astronomy \& Astrophysics, 5071 West Saanich Road, Victoria BC V9E 2E7 Canada}

\author{Tiziana Di Matteo}
\affiliation{McWilliams Center for Cosmology, Department of Physics, Carnegie Mellon University, Pittsburgh, PA, 15213, USA}

\author{Yueying Ni}
\affiliation{McWilliams Center for Cosmology, Department of Physics, Carnegie Mellon University, Pittsburgh, PA, 15213, USA}

\email{kevenr@student.unimelb.edu.au}

% ==== ABSTRACT ==== %
\begin{abstract}

Significant clustering around the rarest luminous quasars is a feature predicted by dark matter theory combined with number density matching arguments. However, this expectation is not reflected by observations of quasars residing in a diverse range of environments. Here, we assess the tension in the diverse clustering of visible $i$-band dropout galaxies around luminous $z\sim6$ quasars. Our approach uses a simple empirical method to derive the median luminosity to halo mass relation, $L_{c}(M_{h})$ for both quasars and galaxies under the assumption of log-normal luminosity scatter, $\Sigma_{Q}$ and $\Sigma_{G}$. We show that higher $\Sigma_{Q}$ reduces the average halo mass hosting a quasar of a given luminosity, thus introducing at least a partial reversion to the mean in the number count distribution of nearby Lyman-Break galaxies. We generate a large sample of mock Hubble Space Telescope fields-of-view centred across rare $z\sim6$ quasars by resampling pencil beams traced through the dark matter component of the BlueTides cosmological simulation. We find that diverse quasar environments are expected for $\Sigma_{Q}>0.4$, consistent with numerous observations and theoretical studies. However, we note that the average number of galaxies around the central quasar is primarily driven by galaxy evolutionary processes in neighbouring halos, as embodied by our parameter $\Sigma_{G}$, instead of a difference in the large scale structure around the central quasar host, embodied by $\Sigma_{Q}$. We conclude that models with $\Sigma_{G}>0.3$ are consistent with current observational constraints on high-z quasars, and that such a value is comparable to the scatter estimated from hydrodynamical simulations of galaxy formation.

\end{abstract}

\section{INTRODUCTION \label{sec:intro}}

The past two decades have yielded significant advances into the search of high redshift quasars, or quasi-stellar objects (QSOs). The collective effort from a number of imaging surveys have classified $\gtrsim 10^{2}$ quasars beyond redshift $z = 6$ (from surveys such as the Sloan Digital Sky Survey, SDSS, \citealt{Richards_2002}; Canada-France High-z Quasar Survey, CFHQS, \citealt{Willott_2010}; Subaru-HSC programmes, e.g. SHELLQs, \citealt{Matsuoka_2018}) including $2$ quasars at $z > 7.5$ \citep{Ba_ados_2017, Yang_2020}, when the universe was a mere fraction of its current age, around $\sim 700$ Myr old. The brightest quasars are some of the most luminous objects observed with bolometric luminosities up to $L_{\rm{bol}} \sim 10^{48}$ ergs$^{-1}$ or equivalently UV magnitudes at $M_{UV} \lesssim -29$. The masses of their central supermassive black holes (SMBH) of order $10^{9} M_{\odot}$ and their relative rarities at $\sim 1$ Gpc$^{-3}$ hint to a picture where these quasars reside within the most massive dark matter halos and are therefore tracers of extreme overdense regions of our Universe \citep{2005Natur.435..629S}. As such, high redshift ($z > 6$) quasars are key objects to an important era of our cosmological timeline in the understanding of early structure formation.

The properties of these quasars are still quite uncertain, with numerous on-going research questions such as their accretion history, seed formation, impact on galaxy formation and SMBH growth mechanisms. The potential to shed light on these processes will therefore have significant impact in constraining models of SMBH growth and galaxy evolution. Fundamentally, these processes are all connected to the halos they are hosted within, yet even in this regard the properties of quasar halo hosts are still ambiguous. On one hand, theory/simulations have typically determined the hosts have masses of the order $\gtrsim 10^{12} M_{\odot}$ at $z \sim 6$, hence are indicative of massive halos forming in rare $(>4 \sigma)$ peaks of the density field based on linear theory \citep{Li_2007, Feng_2014, Costa_2014, Marshall_2019}. These halos are generally expected to be biased with an increased abundance of neighbouring halos in their vicinity, thus one should observe significant cross-correlation between the population of luminous quasars and galaxies in the field-of-view \citep{Overzier_2009, RomanoDiaz_2011, Costa_2014}. 

Observational studies of the cross-correlation of Lyman-break galaxies and Lyman-alpha emitters around luminous quasars at high redshifts have so far yielded no firm conclusion of the local clustering of quasars, with both overdensities \citep{Kim_2009, Utsumi_2010, Husband_2013, Morselli_2014, Garc_a_Vergara_2017, Ota_2018}, and average/underdensities \citep{Stiavelli2005, Willott_2005, Kim_2009, Ba_ados_2013, Mazzucchelli_2017, Kikuta_2017, Uchiyama_2017, He2017, Champagne_2018, Ota_2018} being recorded. One possible factor responsible for the large variance in the environments is the challenge in making direct comparisons of neighbour counts across studies, due to the difference in selection criteria applied on the observations of a limited number of quasar fields. \citet{Willott_2005} using the Gemini telescope finds no evidence of an overdensity of $i$-band dropouts in the quasar field around J1030+0524 relative to the field. However, \citet{Stiavelli2005} observing the same object using the Advanced Camera for Surveys (ACS) from the Hubble Space Telescope (HST) finds a number of missed galaxies fainter than the limiting magnitude used in \citet{Willott_2005} and suggests the quasar field is overdense, containing twice the excess of galaxy dropouts compared to the mean GOODS field. Furthermore, it can appear that the overdensities are apparent on larger scales, beyond the typical field-of-view of a HST/ACS window. The re-observation of $2$ average/underdense \citet{Kim_2009} HST/ACS quasar fields in \citet{Morselli_2014} with a larger field-of-view using the Large Binocular Camera (with a viewing area of $\sim600$ arcmin$^{2}$ compared to HST/ACS's $\sim11$ arcmin$^{2}$) finds that the observed fields are overdense at these larger scales. Additionally, a number of wider area surveys have shown a marked deficit of galaxies around the inner $\sim 2.5$ pMpc region of the quasar, hinting towards the possibility of significant radiative feedback from the AGN that quenches star formation in nearby galaxies \citep{Utsumi_2010, Morselli_2014}. 

In a recent investigation of quasar environments, \citet{Habouzit_2019} use the hydrodynamical simulation Horizon-AGN to infer the distribution of neighbouring galaxies around massive SMBHs, albeit at lower redshifts of $z \lesssim 5$. In their work, they find a high degree of variance for counts for a single projected field-of-view, consistent with observations. Furthermore, an enhancement of number counts relative to the average field should be observable in the $2$cMpc radius around massive SMBHs, although it is contingent on probing to a sufficient sensitivity/depth of field. However, we should also note that the case is more extreme for $z\sim 6$ quasars as the clustering bias is strongly halo mass dependent. We can expect the bias to increase by a factor of $\sim 1.5$ (assuming a \citet{Sheth1999} bias function) between the rarest $10^{-9}$Mpc$^{-3}$ halos versus $10^{-6}$Mpc$^{-3}$ ones available in the Horizon-AGN simulation volume.

Additionally, \citet{Overzier_2009} points out that simply placing the hosts in smaller halos can provide one such interpretation that is consistent with the observed clustering data. In a purely heuristic sense, the median host halo mass for a given quasar luminosity is driven by both the amount of scatter in quasar luminosities and the duty cycle \citep{Veale_2014, Ren_2020}. Conceptually, having scatter facilitates the odds that the brightest quasar/most massive SMBHs are over-luminous outliers hosted within relatively common dark matter halos with lower bias \citep{Lauer_2007}. Similarly, knowledge of the quasar duty cycle, defined here as the proportion of SMBH accreting in the quasar mode at some cosmic time, is also crucial, as any single pointing at a luminous quasar must be during an instance where the quasar is active. Thus a lower duty cycle also infers that quasars are generally contained in more common halos. Both parameters are poorly constrained, as their determination through measurements of clustering and number densities of quasars is challenging due to degeneracies \citep{Wyithe_2009, Conroy_2012, Ren_2020}. 

In this paper, we examine the environments around SDSS-like $z \sim 6$ luminous quasars taking advantage of the modeling framework we developed previously in \citet{Ren_2018} in the context of studying the environment around luminous $z > 8$ galaxies. We simulate mock Hubble Space Telescope (HST) fields-of-view to determine the number of visible Lyman break drop-out galaxies around the most luminous quasars at $z\sim 6$. To achieve this, we utilise the large, high resolution dark matter catalogue underpinning the cosmological hydrodynamical simulation BlueTides, and populate these halos with quasar and galaxy luminosities obtained from our semi-empirical method. Additionally, object to object stochasticity in luminosities is included in the modeling through use of a conditional luminosity function (CLF) prescription. We investigate the available parameter space in both galaxy and quasar scatter and quasar duty cycle that reproduces available measurements of galaxy counts around quasar fields. This paper is outlined as follows. We describe the framework to construct our mock fields-of-view and the model used to populate halos with galaxies and quasars in Section~\ref{sec:method}. In Section~\ref{sec:results} we analyze the diversity of environments, computing the average galaxy number counts visible in a field-of-view for various parameters of scatter. In particular, we evaluate the probability of reproducing the set of HST/ACS observations of \citet{Kim_2009} in the same parameter space. In Section~\ref{sec:conclus}, we conclude with our key findings. For this work, we adopt the standard WMAP9 cosmological parameters \citep{Hinshaw_2013}: $h=0.697$, $\Omega_{m} = 0.2814$, $\Omega_\Lambda = 0.7186$, $\Omega_{b} = 0.0464$, $\sigma_{8} = 0.82$, $n_{s} = 0.971$. All magnitudes are given in the AB magnitude system \citep{Oke_1983}.

\section{Semi-Analytical Setup}  \label{sec:method}

The densities of luminous ($M_{UV} \lesssim -26$) high-$z$ quasars of order $1$Gpc$^{-3}$ play a key role in the difficulty of making any robust statistical inference of their properties. Current state-of-the-art cosmological simulations with volumes comparable to $\sim 1$Gpc$^{3}$, such as hydrodynamical simulation, BlueTides \citep{Feng2015} or the dark matter N-body simulation, Millennium \citep{Springel_2005} are capable of capturing only few of these rare quasars. These simulations are therefore susceptible to small-number statistics in describing the population of these objects. For our model, we utilise an alternative approach and leverage the natural stochasticity present in quasar luminosities residing in halos to simulate a variety of unique environments without the computational cost of running additional N-body simulations. Stochasticity increases the variance for the range of halo masses luminous quasars typically reside in, thus effectively boosting the number of possible environments we can investigate. The small quasar clustering measurements at $z\sim 4$ \citep{He2017} have already hinted at the possibility of significant scatter for luminous quasars, however this possibility is also degenerate with quasars having a low duty cycle, which simply decreases the median halo mass hosting a quasar at fixed luminosity. For our analysis, both the quasar luminosity scatter $\Sigma_{Q}$ and quasar duty cycle $\varepsilon_{DC}$ are kept as free parameters.

Similarly, the luminous galaxies hosted by neighbouring halos are also subject to stochasticity. At minimum, we can consider the variance in the assembly histories for halos to play a non-negligible role in the variation in the star formation rate, and consequently also the galaxy luminosity. Following this, we note that the collection of the most luminous galaxies, while generally expected to reside inside moderately massive halos, can be hosted within more common halos, facilitated by the inherent larger number density. The value of $\Sigma_{G}$ at high redshifts ($z > 6$) is poorly constrained observationally. An earlier study using clustering measurements of $z \gtrsim 8$ galaxies is unable to place limits of $\Sigma_{G}$ owing to insufficient depth in the field-of-view \citep{Ren_2018}. However at lower redshifts, $\Sigma_{G} \sim 0.17 - 0.23$ at $z\sim 0.1$ and is constrained by the bright end shape of the galaxy LF \citep{Yang2003, Cooray_2005}. Simple theoretical modeling finds an approximate constraint for the lower limit, $\Sigma_{G} \sim 0.2$ for $z > 2$ which is inferred from the distribution of halo assembly times \citep{Ren_2018}. Existing modeling of the $z>6$ luminosity function \citep{Ren_2019} and the inferred neutral hydrogen fraction in the intergalactic medium (IGM) at $z\sim 7$ \citep{Whitler_2020} assuming this lower limit of scatter have all yielded consistent results within observation limits. In contrast, simulations can have large inferred ranges between $0.2 < \Sigma_{G} < 0.6$ (additional details provided in Section~\ref{sec:comp}). For brevity, we keep the galaxy luminosity scatter, $\Sigma_{G}$ as a free parameter in this work.
 
Our approach largely follows that of similar work done in \citet{Ren_2018}, investigating the environments of bright $z \sim8$ galaxies. We run a Monte Carlo simulation generating mock fields-of-view centred around the brightest quasars. We use the dark matter halo catalogue of the large cosmological hydrodynamical simulation, BlueTides \citep{Feng2015} and simultaneously populate resolved halos with both a quasar and a galaxy using a conditional luminosity function (CLF) prescription. The key input parameters for the CLF are the object's median luminosity as a function of halo mass, $L_{c}(M_{h})$, log-normal dispersion in the object's luminosity given by a scatter parameter $\Sigma$, and a duty cycle $\varepsilon_{DC}$, for the quasar CLF. 

For each Monte Carlo iteration, we trace multiple pencil beams across the brightest quasars inside the simulation volume with appropriate pencil beam dimensions to emulate the set of observations from \citet{Kim_2009}. To reduce the impact of selection bias, we sample for the top $6$ brightest quasars at every iteration. Galaxies close in projection will be defined as neighbours of the bright object. Here, we assume that only one object is visible per halo. Additionally, the galaxy host of the central quasar is not counted towards the overall number counts. The set of Monte Carlo iterations spans a low and high quasar duty cycle case, $\varepsilon_{DC} = 0.01$ and $1$. The scatter parameter for galaxies and quasars will be systematically probed to both assess the impact on galaxy neighbour counts and to facilitate a comparison with the \citet{Kim_2009} observation set containing 5 quasar fields. The quasars selected in \citet{Kim_2009} have magnitudes between $m_{z} = 19.83 - 20.63$ corresponding to number densities of $\sim 1$Gpc$^{-3}$ and are located at redshifts between $z = 5.99 - 6.40$ (see Section~\ref{sec:comp} for details). We provide additional information on the methods and tools used in the following subsections.

\subsection{Simulation parameters}

For our halo catalogue, we extract the dark matter component from the BlueTides simulation, a state-of-the-art cosmological hydrodynamical simulation for the first galaxies \citep{Feng2015}. BlueTides uses the smoothed particle hydrodynamics (SPH) code, MP-GADGET with $2 \times 7040^3$ particles and tracks their evolution in a cosmological volume of $(400$h$^{-1}$Mpc)$^{3}$ from $z = 99$ to $z = 6.56$. The corresponding mass resolution for the dark matter and gas particles (in the initial condition) in BlueTides are $M_{\rm{dm}} = 1.2\times 10^{7}$h$^{-1}M_{\odot}$ and $M_{\rm{gas}} = 2.4 \times 10^{6}$h$^{-1} M_{\odot}$ respectively. The star particles have a mass of $1/4 M_{\rm{gas}} = 6 \times 10^{5}$h$^{-1}M_{\odot}$. The gravitational softening length is $1.8$ckpc which is indicative of its spatial resolution. Halos are identified in BlueTides with a friends-of-friends algorithm, using a linking length of $b = 0.2$ \citep{Davis_1985}. This configuration results in BlueTides being able to resolve halos down to a mass of $\sim 1 \times 10^9 M_{\odot}$. Finally, AGN feedback in BlueTides is modeled by injecting a fraction of the BH's radiation energy as thermal energy to gas particles in a region twice the radius of the SPH smoothing kernel of the BH particle. The large volume and high mass resolution has enabled BlueTides to conduct detailed studies on the first quasars/most massive black holes \citep{DiMatteo2017, Tenneti_2017, Ni2018, Tenneti_2018} and of the first galaxies \citep{Waters_2016, Waters_2016b, Wilkins_2017}. Additionally, BlueTides has shown to be in full agreement with a number of observables such as the UV galaxy luminosity functions \citep{Feng2015, Wilkins_2017}, galaxy clustering \citep{Bhowmick_2017} and the quasar luminosity function \citep{Marshall_2019, Ni_2020}.

For our analysis, we take the final snapshot available at $z = 6.56$ to compare with the $z \sim 6$ set of quasar observations from \citet{Kim_2009} that vary between $z = 5.99 - 6.40$.  In addition, since added stochasticity in quasars reduces the median halo mass hosting luminous quasars, this suggests that simulations with volumes smaller than $1$Gpc$^{3}$ can be used to sufficiently model statistics around these luminous quasars when given sufficient scatter. We describe in detail the impact of the smaller volume of BlueTides for the environments of quasars in Section~\ref{sec:avg}. 

In each Monte Carlo realisation, pencil beams are traced across the $6$ brightest quasars. Our pencil beams are parameterised with a cross-sectional area $5.92 \times 5.92$ h$^{-2}$Mpc$^{2}$ ($\sim 11.3$ arcmin$^{2}$) and a depth of $244.5$h$^{-1}$Mpc, corresponding to a photometric uncertainty in redshift measurements, $\Delta z = 0.9$ centred at $z=6.56$. We select these physical dimensions to be consistent with the observations of \citet{Kim_2009}, searching for $z\sim 6$, $i$-band dropouts with the HST's Advanced Camera for Surveys (ACS) and a flux limit in the $z$-band of $z_{AB} < 26.5$. Each pencil beam is assigned with a random orientation defined by rotations around two distinct axis. As the size of the simulation volume is larger than the depth of the pencil beam, we have no risk of the pencil beam overlapping with itself.

\subsection{Conditional Luminosity Function}

We utilise an empirical conditional luminosity function (CLF) approach to assign halos with quasar and galaxy luminosities, $L_{Q}$ and $L_{G}$. The CLF, $\Phi(L \mid M_h)$ is a probabilistic description of measuring an object's luminosity, $L$ given some halo mass, $M_{h}$. The CLF for quasars is defined by

\begin{equation}
\Phi(\log L_{Q} \mid M_h)=  (1-\varepsilon_{\rm{DC}}) \delta(L_{Q} = 0) + \dfrac{\varepsilon_{DC}}{\sqrt{2\pi}\Sigma_{Q}}\exp{\bigg( -\dfrac{\log \Big[ {\frac{L_{Q}}{L_{Q,c}(M_{h}, \Sigma_{Q}, \varepsilon_{DC})}} \Big]^{2}}{2\Sigma_{Q}^{2}} \bigg) },
\label{eqn:qclf}
\end{equation}

where $L_{Q,c}(M_{h})$ is the median quasar luminosity as a function of halo mass $M_{h}$, $\Sigma_{Q}$ is the width of the dispersion in dex, $\varepsilon_{DC}$ is a constant quasar duty cycle defined as the fraction of SMBH undergoing quasar-mode accretion and $\delta(x)$ is the Dirac-delta function. Similarly, we define the CLF for galaxies as

\begin{equation}
\Phi(\log L_{G} \mid M_h)=\dfrac{1}{\sqrt{2\pi}\Sigma_{G}}\exp{\bigg( -\dfrac{\log \Big[ {\frac{L_{G}}{L_{G,c}(M_{h}, \Sigma_{G})}} \Big]^{2}}{2\Sigma_{G}^{2}} \bigg) },
\label{eqn:gclf}
\end{equation}

with $L_{G,c}(M_{h})$, $\Sigma_{G}$ as the galaxy-equivalent variables for $L_{Q,c}(M_{h})$ and $\Sigma_{Q}$. We assume a log-normal form for our CLF as standard in literature for defining galaxy observables as a function of halo mass (Luminosities: \citealt{Cooray_2005, Yang_2009}. Stellar masses: \citealt{Behroozi2010, Moster_2010, Reddick_2013}). The assumption of a log-normal functional form (to first order at least) is easily justified by invoking the multiplicative central limit theorem for the collective processes that link halo mass to luminosities. Additionally, the CLF is related to the LF by

\begin{equation}
\phi(\log L) = \int^{\infty}_{0} \dfrac{dn}{dM_{\rm{h}}} \Phi(\log L \mid M_{\rm{h}}) dM_{\rm{h}},
\label{eqn:lf}
\end{equation}

where $ \frac{dn}{dM_{\rm{h}}}$ is the BlueTides halo mass function. We follow the simple procedure outlined in \citet{Ren_2019} to derive the median luminosity versus halo mass relation, $L_{c}(M_{h}, \Sigma)$ with enfolded stochasticity. The \citet{Ren_2019} procedure in short, is based on modifying $L_{c}(M_{h}, \Sigma = 0)$, i.e. standard abundance matching without scatter, by a constant scaling factor and substituting a constant luminosity value past a characteristic threshold mass, $L_{c}(M_h > M^{c}_{h}) = L_{c}(M^{c}_{h})$ to achieve a well-fitting modeled LF through using Equation~\ref{eqn:lf}. The physical interpretation of $M^{c}_{h}$, can be linked to the accretion of hot quasi-static gas around the host galaxy onto its SMBH, i.e. radio-mode AGN feedback, which is expected to be substantial for massive galaxies \citep{Croton2006}. For the abundance matching, we use the \citet{Bouwens2015} $z \sim 7$ galaxy LF\footnote{
The choice of \citet{Bouwens2015} galaxy LF results in a higher number of $m_{z} < 26.5$ galaxies compared to other $z\sim7$ galaxy LFs from \citet{Finkelstein2015} and \citet{Bowler2016} by $\gtrsim 30\%$. Thus, using the \citet{Bouwens2015} LF remains a more conservative case as it increases the likelihood of finding overdense environments.} and the \citet{Matsuoka_2018} $z\sim6$ quasar LF, that are subsequently redshift-evolved to $z=6.56$ to match the redshift of the BlueTides snapshot. Heuristically, simply using $\Sigma > 0$ with the zero-scatter relation, $L_{c}(M_{h}, \Sigma = 0)$ will overestimate the bright end of the LF \citep{Cooray_2005}. Thus, in \citet{Ren_2020}, we demonstrate that using a simple mass cut-off, $M^{c}_{h}$ offers a good approximation over standard deconvolution techniques for modeling the bright end of the LF under the influence of non-zero scatter. More importantly, the cut-off $M^{c}_{h}$, has an intuitive interpretation for feedback in $L_{c}(M_{h})$ that evolves with $\Sigma$, where a higher $\Sigma$ requires self-regulation to occur at lower halo masses in order to preserve the LF. This is a feature may otherwise be hidden due to `overfitting' from using a deconvolution method. The scaling factor and characteristic threshold parameters are calculated by chi square minimization to measurements of the LF and the resulting $L_{c}(M_{h})$ is then normalized to the average number of galaxies detected in a random pointing at $\Sigma = 0$. As our limiting constraint is in the $z$-band, we convert our $1450${\AA} magnitudes to $z$-band magnitudes by calculating the UV slopes, $\beta$, and assuming the galaxy UV continuum, $f_{\lambda} \propto \lambda^{\beta}$ where $\beta = -2 - 0.2(19.5 - M_{UV})$ \citep{Bouwens_2014}.

We take a conservative lower limit of galaxy luminosity scatter, $\Sigma_{G} = 0.1$ based on the analytical estimate of $\Sigma_{G} \gtrsim 0.2$ from halo assembly processes \citep{Ren_2018}. Similarly, we tentatively select $\Sigma_{G} = 0.5$ as our upper limit, guided by the average $\langle \Sigma_{G} \rangle \gtrsim 0.4$ from Meraxes \citep{Ren_2018}. The lower limit of the quasar luminosity scatter, $\Sigma_{Q} = 0.4$ is used to allow for a greater range of halo masses when selecting for luminous quasars, thus reducing the impact of selection bias. We find that $\Sigma_{Q} \lesssim 0.3$ does not achieve enough variety in halo masses with our catalogue, leading to a skewed distribution of galaxy neighbours for the $\Sigma_{Q} = 0.3$ case due to the presence of a lower density environment around a particularly massive halo. We note that the effect of selection bias can be reduced using a larger simulation box or taking the aggregate result of multiple viewings. We adopt an upper limit of $\Sigma_{Q} = 0.8$, as it becomes challenging to determine $L_{Q,c}(M_{h})$ that adequately fits the knee in the \citet{Matsuoka_2018} QLF for values that are significantly greater $\Sigma_{Q} > 0.7$. Thus, we vary our free parameters over the range, $\Sigma_{G} \in [0.1,0.5]$ and $\Sigma_{Q} \in [0.4, 0.8]$, while assuming values $\varepsilon_{DC} = 0.01$ and $1$ for quasars. In Fig.~\ref{fig:fig1}, we show the derived median luminosity against halo mass relations, $L_{c}(M_{h})$ with the limiting values of $\Sigma$ for each object type. The resulting modeled LFs computed from $L_{c}(M_{h})$ using Equation~\ref{eqn:lf} are consistent with observations within uncertainties.

\section{Results} \label{sec:results}

\subsection{The diversity of environments around luminous quasars} \label{sec:diver}

Our model samples over the parameter space $\Sigma_{G} \in [0.1,0.5]$, $\Sigma_{Q} \in [0.4,0.8]$ and considers two duty cycles, $\varepsilon_{DC} = 0.01$ and $1$. For each point in the $(\Sigma_{G},\Sigma_{Q}, \varepsilon_{DC})$ parameter space, we run a total of $\sim 10^{3}$ Monte Carlo realizations. We follow the \citet{Kim_2009} observation parameters, using a magnitude limit $m_{z} < 26.5$ with a pencil beam cross-sectional area matching the HSC/ACS field-of-view $11.3$ arcmin$^2$. We collect the number of visible galaxies centred over the top $6$ most luminous quasars in each Monte Carlo realization to reduce the impact of selection bias. With the targeted fields, we follow the same definition from \citet{Kim_2009}, where the number of galaxies visible in the quasar field does not include the host galaxy. In addition for every targeted quasar pointing, we also collect the number of visible galaxies from a random field pointing in the same simulation volume, i.e. we collect $6$ random fields per Monte Carlo realization. The average number of visible galaxies per pointing in the field throughout the simulation volume is $\sim 3.7 \pm 2.6$ galaxies, calculated by averaging random pointings across all $\Sigma_{G}$. This value is consistent with the number of galaxies expected in a GOODS ACS field, between $3-8$ galaxies \citep{Kim_2009}. We find that the average number of visible galaxies per pointing is insensitive to $\Sigma_{G}$, which is expected as we calibrate our LF at every $\Sigma_{G}$. However, we find a small increase in the dispersion of galaxy counts from the random field pointings with $\Sigma_{G}$, from $2.3$ galaxies with $\Sigma_{G} = 0.1$ to $2.9$ galaxies with $\Sigma_{G} = 0.5$.

In Fig.~\ref{fig:fig0}, we show select snapshots demonstrating the diversity in visible galaxy neighbours around highly luminous quasars using our simple empirical approach (bottom panels). In addition, we also include analogs to our quasar fields as predicted by the full hydrodynamical suite BlueTides for comparison (top panels; see \citealt{Feng2015, DiMatteo2017, Marshall_2019} for details on the sub-grid prescription used in BlueTides). For the BlueTides sample, we similarly trace pencil beams across the simulation volume and record halos captured inside the beam. We compute the galaxy luminosity by summing the luminosities from all star particles associated to a halo. The UV luminosity from each star particle was derived using the Binary Population And Spectral Population Synthesis models (BPASS v2.2, \citealt{Stanway_2018}) with the particle mass, age and metalicity as inputs \citep{Wilkins_2017}. We apply the photometric correction in a similar fashion to our empirical modeling, converting the magnitude from UV to the $z$-band. Our approach does not distinguish between satellite galaxies as their luminosity contribution to the overall luminosity is small (one can estimate this contribution using the stellar masses of $z=7.5$ satellites and centrals from BlueTides; \citealt{Bhowmick_2018}). Finally, we use the same magnitude limit of $m_{z} < 26.5$ to select for our galaxy neighbours. In BlueTides, we note a diversity of environments selected from the sample of the top $6$ most massive SMBHs, with the neighbour count ranging from $8 - 36$ galaxies (3 of which are shown in Fig.~\ref{fig:fig0}) compared to the baseline value of $11.4 \pm 6.2 $ galaxies from $2 \times 10^{3}$ random pointing pencil beams. We note that the results given here are not corrected for dust, thus both the number counts of neighbours as well as magnitudes for the objects can be potentially overestimated \citep{Marshall_2019, Ni_2020}. Despite not being able to determine the extent of dust attenuation in our BlueTides quasar fields, the panels in Fig.~\ref{fig:fig0} still qualitatively show the diversity of environments with respect to both halo mass and central quasar luminosity. In BlueTides, there is a large variation between halo mass and quasar luminosities which arrives as a consequence of SMBH masses being correlated with a low-tidal environment \citep{DiMatteo2017, Huang_2020, Ni_2020b}. Phenomenologically, low-tidal fields have the capability to induce rapid growth of the central SMBH through direct radial accretion of cold gas from filaments. This correlation facilitates the large variation in quasar luminosities, and subsequently can lead to a variety of densities around luminous quasar fields.

We also check for any correlation between quasar luminosity and the environment in BlueTides by individually comparing the set of $9$ most luminous quasars to other lower luminosity quasars with similar halo masses. We define lower luminosity quasars as being fainter by at least $3$ mag to our target quasar, and take similar mass halos from the range $\Delta \log(M_{h}) = 0.1$ of the target. We account for projection variation by simulating $1\times10^{2}$ randomly oriented pencil beams around each target luminous AGN candidates. We find that this variation can be significant having a dispersion up to $50\%$ of the average neighbour count around the quasar. Keeping in mind that AGN feedback in BlueTides not modeled by radiative transport, meaning it does not capture possible effects of quasar radiation across Mpc scales onto nearby central galaxies. Thus, any correlation, if any would be an indication that the conditions that enable luminous quasars also facilitate star formation in nearby galaxies.

In Fig.~\ref{fig:fig05} we show reference single random pointings around the $9$ most luminous quasars, together with the distribution of the number of galaxy neighbours from the set of fainter quasars. We find that the quasar luminosity is not significantly correlated to the number of visible galaxy neighbours in the field of view. For example, as shown in the upper left panel, the set of fainter quasars inside similar mass halos to the most luminous quasar are considered to be overdense compared to the field average with a count of $\sim 20.0 \pm 8.3$ galaxies to $11.4$ galaxies. However, the most luminous quasar field containing $24.9 \pm 8.8$ visible galaxy neighbours is well within $1\sigma$ of the sample average of the set of fainter AGN counterparts.

\begin{figure}[ht!]
	\centerline{\includegraphics[angle=-00, scale=0.84]{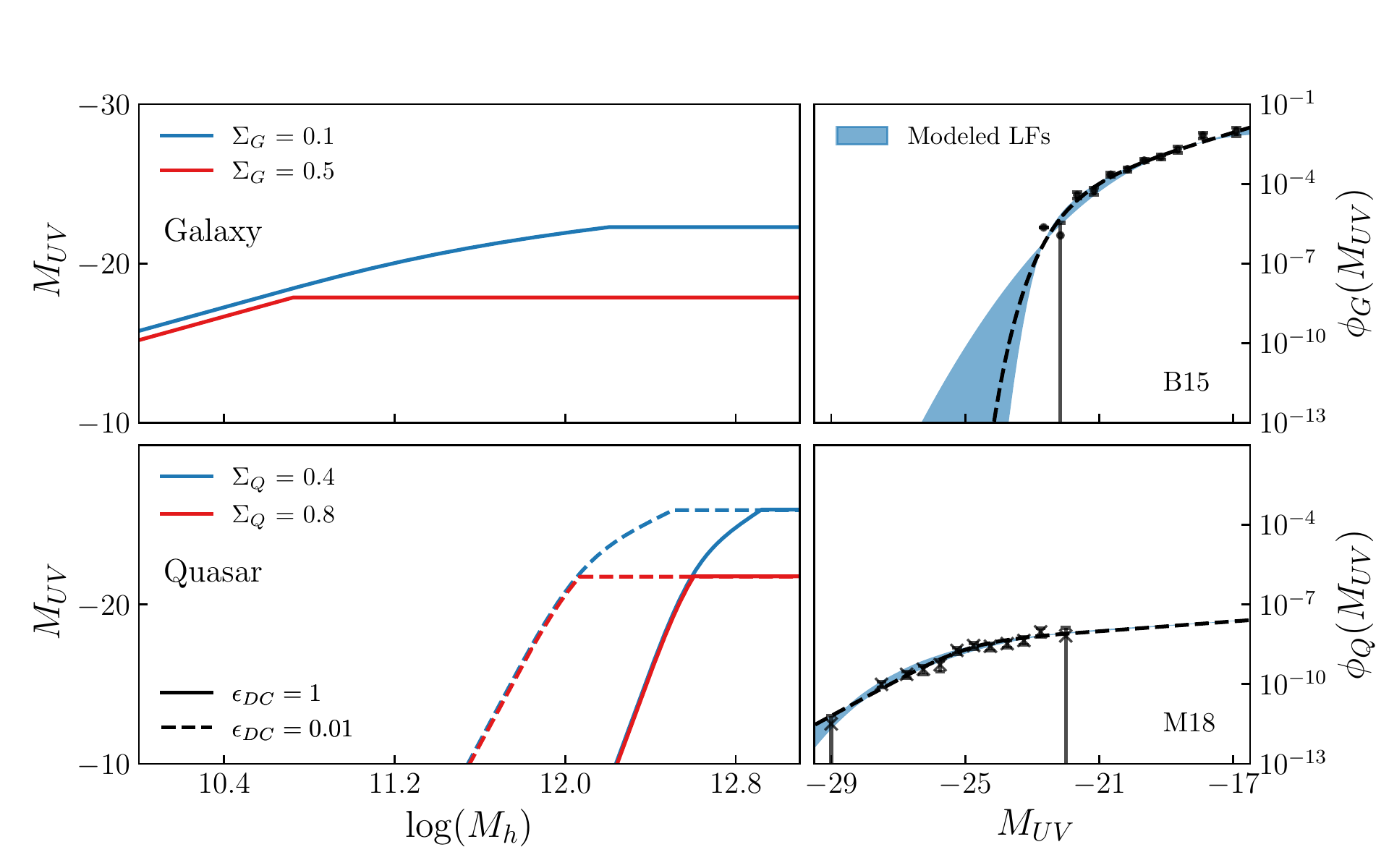}}
	\caption{\small The modeled median luminosity versus halo mass relation, $L_{c}(M_{h})$ for: galaxies (upper left) and quasars (bottom left), assuming minimum (blue curves) and maximum (red curves) scatter, $\Sigma$ cases. For quasars we also vary the duty cycle, $\varepsilon_{DC} = 0.01, 1$ (dashed and solid). $L_{c}(M_{h})$ is calibrated to the luminosity function, $\phi(M_{UV})$ (right) by iteratively solving for $\phi(M_{UV})$ in Equation~\ref{eqn:lf}. The range of calibrated luminosity functions across $\Sigma$ is given by the shaded region: galaxies (top right) and quasars (bottom right). The solid black points and dashed black curves are the observational data points plus best fits lines respectively for \citet{Bouwens2015} (galaxies, upper right) and \citet{Matsuoka_2018} (quasars, lower right). Note that some error bars in the galaxy luminosity function are smaller than their data points. }
	\label{fig:fig1}
\end{figure}

For the demonstration of our empirical model, we select the values $\Sigma_{G} = 0.3$, $\Sigma_{Q} = 0.5$ and $\varepsilon_{DC} = 1$. The different densities shown in Fig.~\ref{fig:fig0} are relative to the bright quasar pointings as opposed to the field, with the lower central panel in Fig.~\ref{fig:fig0} representing the average environment of a bright quasar field. Both BlueTides and our empirical model are qualitatively consistent in that the majority of luminous quasars are typically hosted by massive halos and intrinsically lie in highly biased regions, but also having a fewer number of luminous quasars in less dense environments. For the selected set of parameters, the average number of visible $m_{z} < 26.5$ galaxy neighbours detected in a quasar field-of-view is $\sim 9.1$ galaxies, however the variation is large with a standard deviation of $\sim 3.9$ galaxies per field-of-view. Under the most extreme parameter values $\Sigma_{G} = 0.5$ and $\Sigma_{Q} = 0.8$, the average number falls to $\sim 2.9$ galaxies (excluding the galaxy in the central halo) with a standard deviation of $\sim 1.9$ galaxies. The trend of the variance in visible galaxy counts exceeding the Poissonian limit is not unexpected due to cosmic variance \citep{Trenti_2008}, but here also finds another source in the scatter, $\Sigma$. In fact, we show that a great diversity of environments can be easily achieved through the inclusion of some scatter. We find that the characterization of weak clustering can be a consequence of $\Sigma_{G}$, which is seen by comparing both the underdense and the average density panels in Fig.~\ref{fig:fig1}. Both these panels have their central quasars occupying the same mass halo embeded within a highly overdense region with halo bias, $b \sim 2 \times 10^{2}$. On the contrary, the overdense panel shows a quasar residing inside a comparatively less massive host to the previous two cases, but shares an overabundance of visible galaxies in its field-of-view.  At face value, we can infer that the distribution of the visible neighbours can be dominated by the processes that facilitate galaxy evolution (i.e. processes that directly contribute to the value of $\Sigma_{G}$ such as, the variation in star formation rates; \citealt{Ren_2018}, or the scatter in stellar masses for a given halo mass; \citealt{Behroozi2010}) in a halo rather than the advantage in halo counts originating from large scale structure phenomenology. 

Additionally, we highlight in Fig.~\ref{fig:fig1} that $\Sigma_{G}$ not only increases the variation of galaxy luminosities at a given halo mass but also sets the scale for radio-mode AGN feedback, as indicated by the mass where $L_{c}(M_{h})$ flattens. For example, $\Sigma_{G} = 0.3$ points to a feedback scale corresponding to $M_{h} \sim 10^{11.5} M_{\odot}$, in general agreement with the peak star formation efficiency which occurs at $M_{h} \sim 10^{12} M_{\odot}$, at high redshift \citep{Tacchella2018, Behroozi2019}. Seen this way, $\Sigma_{G}$ offers two distinct mechanisms to explain an underdense observation: (1) large variation in galaxy luminosities increases the variance of visible galaxy number counts in the field, boosting the odds of a serendipitous underdensity observation and (2) neighbouring galaxies are increasingly likely to be self-regulated, especially at high $M_{h} \gtrsim 10^{11.5}M_{\odot}$.

This result eases tension in models finding strong clustering conditions around quasars with assumed $\varepsilon_{DC} \sim 1$. These are typical outputs from using simpler models for populating halos with galaxies, such as assigning Lyman-break galaxies through a constant halo mass cut \citep{RomanoDiaz_2011, Buchner_2019}. In these instances, the use of a halo mass cut does not consider the event where the massive surrounding halos may be self-regulated, thus leading to an overestimation of visible galaxy neighbours.

\subsection{Average number of galaxy neighbours in the QSO field} \label{sec:avg}

In Fig.~\ref{fig:dcboth} we show the average number of galaxy neighbours visible in a single bright quasar field as a function of our free parameters $\Sigma_{G}$, $\Sigma_{Q}$ and $\varepsilon_{DC}$. Perhaps unremarkably, the average bright quasar field tends to be in excess of neighbours compared to a random field pointing ($\sim 3.7$ galaxies) ranging from an average of ($\sim 4 - 20$) neighbours for $\varepsilon_{DC} = 1$ and ($\sim 3 - 12$) neighbours for $\varepsilon_{DC} = 0.01$. Here, the highest neighbour counts occur at the lowest $\Sigma_{G}$ values. However, the variance in each of these fields can also be significant as mentioned in Section~\ref{sec:diver}.

The two most prominent parameters that determine the extent of clustering are $\Sigma_{G}$ and $\varepsilon_{DC}$. It is relatively intuitive to understand how either of these parameters should impact clustering around quasars. As $L_{c,G}(M_{h}, \Sigma_{G})$ is calibrated to the LF, having $\Sigma_{G} > 0$ effectively redistributes the population of galaxies at a fixed luminosity across an increasing range of halo masses. This decreases the probability for massive haloes with similarly massive neighbours to contain luminous galaxies and reduces the overall number of galaxies in the average quasar field-of-view. On the contrary, the duty cycle operates independently to $\Sigma_{G}$ by assuming a constant probability, $\varepsilon_{DC}$ for a halo to contain an active SMBH. This reduces the nominal halo mass for the most luminous quasars implying generally less clustered environments. Finally, the effect of the quasar scatter, $\Sigma_{Q}$ should be degenerate with $\varepsilon_{DC}$ where both of these parameters would alter the distribution of host halo masses for quasars of fixed luminosity. 

Curiously, the average galaxy number counts centred on luminous quasars appears to only show weak to no dependence in $\Sigma_{Q}$ across the entire parameter space. We explore this discrepancy by deriving the distribution of the linear bias factor for the halos containing our brightest sources in the simulation,

\begin{equation}
p(b_{\mathrm{brightest}}) \propto \int \dfrac{dn}{dM_{\rm{h}}}(M_{\rm{h}}(b)) \Phi(\log L \mid M_{\rm{h}}) p(L_{\mathrm{brightest}}) d\log L,
\label{eqn:bias}
\end{equation}

where $M_{h}(b)$ is the inversion of the \citet{Sheth1999} halo-mass bias relation, $ \Phi(\log L \mid M_{\rm{h}})$ is our usual conditional luminosity function and $p(L_{\mathrm{brightest}})$ is the distribution of luminosities for the brightest object given by the following relation,

\begin{equation}
 p(L_{\rm{brightest}}) \propto \dfrac{d}{dL_{\rm{brightest}}} \Bigg( \int_{L_{\rm{min}}}^{L_{\rm{brightest}}} \phi(L) dL \Bigg )^{n(V)}.
\label{eqn:mag}
\end{equation}

Here, $\phi(L)$ is the quasar luminosity function (QLF), $n(V)$ is the expected number of objects in a volume $V$ between the range of magnitudes ($L_{\rm{min}}$, $L_{\rm{max}}$),

\begin{equation}
n(V) \approx  \int_{L_{\rm{min}}}^{L_{\rm{max}}} V \phi(L) dL.
\label{eqn:n}
\end{equation}

We take our limits ($L_{\rm{min}}$, $L_{\rm{max}}$) to be $M_{UV} = (-20,-30)$ respectively. 

In Fig.~\ref{fig:bias} we show the probability distribution for the bias of halos hosting our brightest quasars at our limiting $\Sigma_{Q}$, values and the probability distribution for the magnitude of the brightest object in a cosmic volume, $V$. The duty cycle is taken to be $\varepsilon_{DC} = 1$, as changes in duty cycle do not have any significant impact on the spread of bias for a quasar at a given luminosity \citep{Ren_2020}. Note that the sharp peak is an artifact from our modeling method by assuming a sudden cutoff in the median quasar luminosity versus halo mass relation (Fig.~\ref{fig:fig1}). We find that for a simulation volume similar to that of BlueTides ($400$h$^{-1}$Mpc)$^{3}$, our analysis should measure only a small dependence on $\Sigma_{Q}$ for clustering around the brightest quasar. In the parameter space where $\Sigma_{Q}$ is low, the brightest quasar should still have significant probability of being hosted inside halos with $M_{h} < M_{h}^{c}$, where $M_{h}^{c}$ is the characteristic cut-off in $L_{c}(M_{h}, \Sigma_{Q})$. Thus, the instance where we do not see tangible differences between $\Sigma_{Q}$ values is just a limitation of having an insufficient simulation volume to fully capture $M_{h} > M_{h}^{c}$ halos. Additionally, we would also expect any measured trends to be further diluted as our analysis uses the average of the visible galaxy counts around the top $6$ most luminous quasars rather than the single brightest quasar. We also show the expected distribution of bias around the brightest quasar assuming sufficient volume in the bottom panel of Fig.~\ref{fig:bias}. We do not expect moving to a larger simulation to have any noticeable qualitative impact given the overlap in the bias distribution even with a volume such as ($1.5$Gpc)$^{3}$. However, if we were to still repeat our Monte Carlo analysis using the larger volume ($1.5$Gpc)$^{3}$, then it would be easier to infer a small dependency in $\Sigma_{Q}$ for clustering around the brightest quasars compared to this analysis using a volume of ($400$h$^{-1}$Mpc)$^{3}$.

Another interesting aspect from Fig.~\ref{fig:bias} is that variance in the bias, or the diversity of environments around the brightest quasar is larger for smaller $\Sigma_{Q}$. Intuitively, we would expect the opposite effect as a larger $\Sigma_{Q}$ implies a wider range of luminosities at a given $M_{h}$. The rationale is that this effect is offset by the combination of a decrease in flattening threshold mass $M_{h}^{c}$ as $\Sigma_{Q}$ increases in combination with the strong dependence on $M_{h}$ for the bias. Additionally, the figure also shows how $\Sigma_{Q}$ impacts the luminosity dependent nature of clustering. We see that the distribution of bias for the brightest object does not notably change even as volume is increased for the highest values of $\Sigma_{Q}$. This implies that the clustering around quasars brighter than the quasar luminosity $L$, where $\phi_{Q}(L) = 1/V^{*}$ is insensitive to luminosity. Here, $V^{*}$ is defined as the smallest volume such that $p(\rm{bias}_{\rm{brightest}})$ does not change with increasing $V$. Conversely, we see an evolution in the bias around the brightest quasars for low values of $\Sigma_{Q}$ when increasing volume. This indicates that there is some degree of luminosity-based clustering until we reach this limiting luminosity. We can understand this in the context of our introduced flattening from the radio-mode AGN feedback in our modeling of the median luminosity versus halo mass relation, where the median luminosity is no longer dependent on halo mass past $M_{h} > M_{h}^{c}$.

\subsection{Comparison with the observations of Kim et al.} \label{sec:comp}

\citet{Kim_2009} presents a sample of $5$ quasar fields at $z \sim 6$, including a single field from \citet{Stiavelli2005}, analyzing the number of $i$-band dropout galaxies (limiting magnitude of $z_{AB} < 26.5$) in the vicinity of the central quasar. The redshift range of the quasars spans from $z = 5.99 - 6.40$, slightly lower than the redshift $z = 6.56$ used in this study. The magnitude of the central quasar ranges from $z_{850} = 19.98 - 20.63$ corresponding to a number density of $\sim 10^{-9}$Mpc$^{-3}$. The quasar fields examined by the \citet{Kim_2009} study finds a large variance in galaxy number counts, yielding in total $2$ underdense fields (relative to the normalized average number count in the GOODS survey), $2$ overdense fields and $1$ with an average density. A number of simulations have attempted to derive a theoretical understanding behind the variation in these number counts around massive SMBH at high-z \citep{RomanoDiaz_2011, Costa_2014, Habouzit_2019}. A common thread between these simulations is that the environments around the most massive SMBHs are overdense relative to the average counts in the field. Critically, this finding persists even with a smaller scale simulation ($100$Mpc)$^{3}$ at lower redshift $z \sim 5$, implying that clustering should be further enhanced with larger masses and at higher $z$. Additionally, both \citet{Costa_2014} and \citet{Habouzit_2019} have also reported significant variations in the fields centred around massive SMBHs. In Fig.~\ref{fig:dcboth} and Fig.~\ref{fig:bias} we show that the modeling described here is qualitatively consistent with the results of these simulations in both density and variance of fields. 

While these results have generally found that the local neighbourhoods of massive SMBHs tend to be clustered, we arrive at a source of tension for the multiple observational accounts of underdense fields around these luminous quasars. Various attributed physical explanations can include: having a variable galaxy duty cycle \citep{RomanoDiaz_2011}, strong galactic winds from supernovae \citep{Costa_2014}, or radiative, quasar-mode feedback from surrounding AGNs across megaparsec scales \citep{Habouzit_2019}. Another plausible explanation is that the \citet{Kim_2009} fields are too shallow and the Poisson noise becomes comparable to the signal, hence deeper imaging may be required in order to robustly claim a lack of clustering signal for these underdense fields. However, there is still scope in using these shallow number counts to inform the range of our scatter parameters $\Sigma_{Q}$, $\Sigma_{G}$ and to also see if scatter presents a viable explanation in relieving the tension in these findings.

We assess this tension by thoroughly investigating the possibility of claiming multiple detections of $>3$ overdense fields (equivalent to the complement of finding $>2$ underdense fields) over a quintuple set of images. For brevity, we remove the average density classification and have labeled the fields as either overdense or underdense. Here, overdense is defined here as having a galaxy number counts in excess compared to the average random pointing of $\sim 3.7$ galaxies (i.e. $4$ or more galaxies), while underdense is simply defined as the inverse. We group the individual quasar pointings into sets of $5$ and derive the probability of replicating the \citet{Kim_2009} sample after assuming values of $\Sigma_{G}$ and $\Sigma_{Q}$. On a fundamental level, this probability depends on both the average number and the variance in galaxy number counts in the quasar pointings. This Monte Carlo approach complements previous studies of investigating the environments by having the capacity to generate a sample size large enough for statistical inferences. In Fig.~\ref{fig:dcboth2ud5} we show the confidence level of obtaining $>3$ overdense fields out of a set of $5$. As expected, the tension in having multiple underdense fields in a set of $5$ fields can be high. With our nominal assumption of maximal duty cycle, $\varepsilon_{DC} = 1$, the 2$\sigma$ contour (corresponding to a $\sim 5\%$ probability of matching \citet{Kim_2009} results) is restricted to the curve $\Sigma_{G} \sim 0.3$ and lies relatively independent on $\Sigma_{Q}$. On the contrary, a lower duty cycle, $\varepsilon_{DC} = 0.01$ opens up the parameter space and implies that it is possible that such a scenario is plausible for all values of $\Sigma_{G}$.

Similar to the results of Fig.~\ref{fig:dcboth}, the most prominent parameters that determine the likelihood of replicating the observations of \citet{Kim_2009} are $\Sigma_{G}$ and $\varepsilon_{DC}$. While our model shows that setting a low duty cycle presents a suitable explanation, the determination of $\varepsilon_{DC}$ at high redshifts is poorly constrained in practice and remains an open question of research. In an heuristic sense, a low duty cycle is theoretically constrained by observations of $10^{9} M_{\odot}$ SMBHs by $z \sim 7$ which effectively places further constraints on the formation modes of seeds and early accretion processes. For example, hydrodynamical simulations (BlueTides, Appendix~\ref{apdx:a}; Illustris, \citealt{DeGraf_2016}) and theoretical modeling \citep{Aversa2015} have inferred duty cycles of order unity for $z > 6$ SMBHs. However, observational determinations of the duty cycle at high redshifts $z > 4$, typically using quasar-galaxy cross-correlation measurements have alluded to a wide range of duty cycles between $1 \times 10^{-3} < \varepsilon_{DC} < 1$ \citep{Shen2007, Shankar2010, Shankar_2010, He2017}. Addressing these tensions on $\varepsilon_{DC}$ is beyond the scope of this study. However we will note that derivations of the duty cycle based on luminosity measurements can be highly sensitive to obscuration effects from both the quasar and galaxy \citep{Chen_2018, Trebitsch_2019} or even simply from the definition of the minimum luminosity that constitutes an `on' quasar \citep{DeGraf_2016}. In a similar way, the measurements of the duty cycle can also be potentially underestimated if there is a population of feedback-affected galaxies that are unseen, but exist at fainter magnitudes as detailed in Section~\ref{sec:avg}. Taking the uncertainty in low duty cycle scenarios, we investigate if a cohesive picture can be built upon the worst-case assumption of maximal clustering with duty cycle of order unity. 

The natural question would be to enquire to the current constraints in $\Sigma_{G}$. An estimated theoretical lower limit at $z > 6$ is $\Sigma_{G} \sim 0.2$ \citep{Ren_2018} derived from the variance in dark matter halo assembly times. More sophisticated modeling has yielded various estimates of galaxy luminosity scatter at $z > 6$, all higher than the derived lower limit: $\Sigma_{G} \sim 0.32 $ (BlueTides, hydrodynamical, Appendix~\ref{apdx:b})), $\sim 0.38 - 0.58$ (Meraxes, semi-analytical, \citealt{Mutch_2016}), $>0.2$ ($z$-independent galaxy evolution model, empirical, \citealt{Tacchella2018}), $ > 0.25$ (UniverseMachine, empirical, \citealt{Behroozi2019}), $ > 0.3$ (IllustrisTNG, hydrodynamical, \citealt{Vogelsberger_2020}). The latter three constraints are considered lower limits in $\Sigma_{G}$ as they are the scatter derived from the stellar mass to halo mass relation and UV luminosity to stellar mass relation respectively. Here, we just assume the existence of some additional processes that links their listed scatter to $\Sigma_{G}$ to be summed in quadrature. We can see that on face value, most of the galaxy scatter measured by simulations have largely not excluded the \citet{Kim_2009} observations to a $2\sigma$ level. However, it is also worthwhile to note that our result shows a surprising sensitivity in $\Sigma_{G}$ to the likelihood of finding multiple underdense fields. In the nominal case, $\varepsilon_{DC} = 1$ we see that only a minor boost of $\Delta \Sigma_{G} \sim 0.05$ from $\Sigma_{G} = 0.3$ to $0.35$ is needed for a significant increase in overall probability $\sim 5\%$ to $\sim 32\%$. While tight, hydrodynamical simulations tend to hover close around the value $\Sigma_{G} > 0.3$, hence it is possible that any observed underdensities in terms of galaxy neighbours falls completely within expectations.

\section{Conclusion \& general remarks} \label{sec:conclus}

In this paper, we revisit the existing tension in the diversity of environments around $z \sim 6$ quasars between theory/simulations predicting high clustering and observations with measurements of weak/average clustering. We approach this problem with an empirical method, populating halos in a high resolution N-body dark matter only simulation from BlueTides at $z = 6.56$ with quasar and galaxy luminosities. The simulation's box volume of ($400\mathrm{h}^{-1}$Mpc)$^{3}$ is sufficiently large to track the rare host halos that could plausibly host SDSS-like quasars. The relations between halo mass and the object's luminosity are calibrated to their luminosity functions. We then create mock field-of-view images by tracing pencil beams through luminous quasars and recording the number of visible galaxies within a $5.92 \times 5.92$h$^{-2}$Mpc$^{2}$ area plus a depth corresponding to the photometric uncertainty $\Delta z \sim 0.9$, at $244.5$h$^{-1}$Mpc. Specifically, we explore the possibility of scatter when populating galaxy or quasar luminosities ($\Sigma_{G}$, $\Sigma_{Q}$) as a source to alleviate tensions between modeling and the specific observations of \citet{Kim_2009}. In addition to scatter, we also investigate the impact of different values of a constant quasar duty cycle, $\varepsilon_{DC}$, defined here as the relative proportion of black holes actively undergoing quasar-mode accretion.

The innovation of our work comes from leveraging stochasticity to generate a large sample of distinct fields. Using a Monte Carlo method, we resample galaxy and quasar luminosities in our simulation volume over $\gtrsim 10^{3}$ times for every point in parameter space. Scatter reduces the average halo mass hosting an object of given luminosity by increasing the probability for common halos to accommodate a quasar with an outlier accretion rate, thus the number of independent halo conditions accessed increases exponentially for the rarest brightest objects. This method allows rapid exploration of the parameter space ($\Sigma_{G}, \Sigma_{Q}$) that is relatively statistically robust at minimal cost, hence maintaining a competitive advantage over traditional modeling methods.

\begin{figure}[ht!]
	\centerline{\includegraphics[angle=-00, scale=0.60]{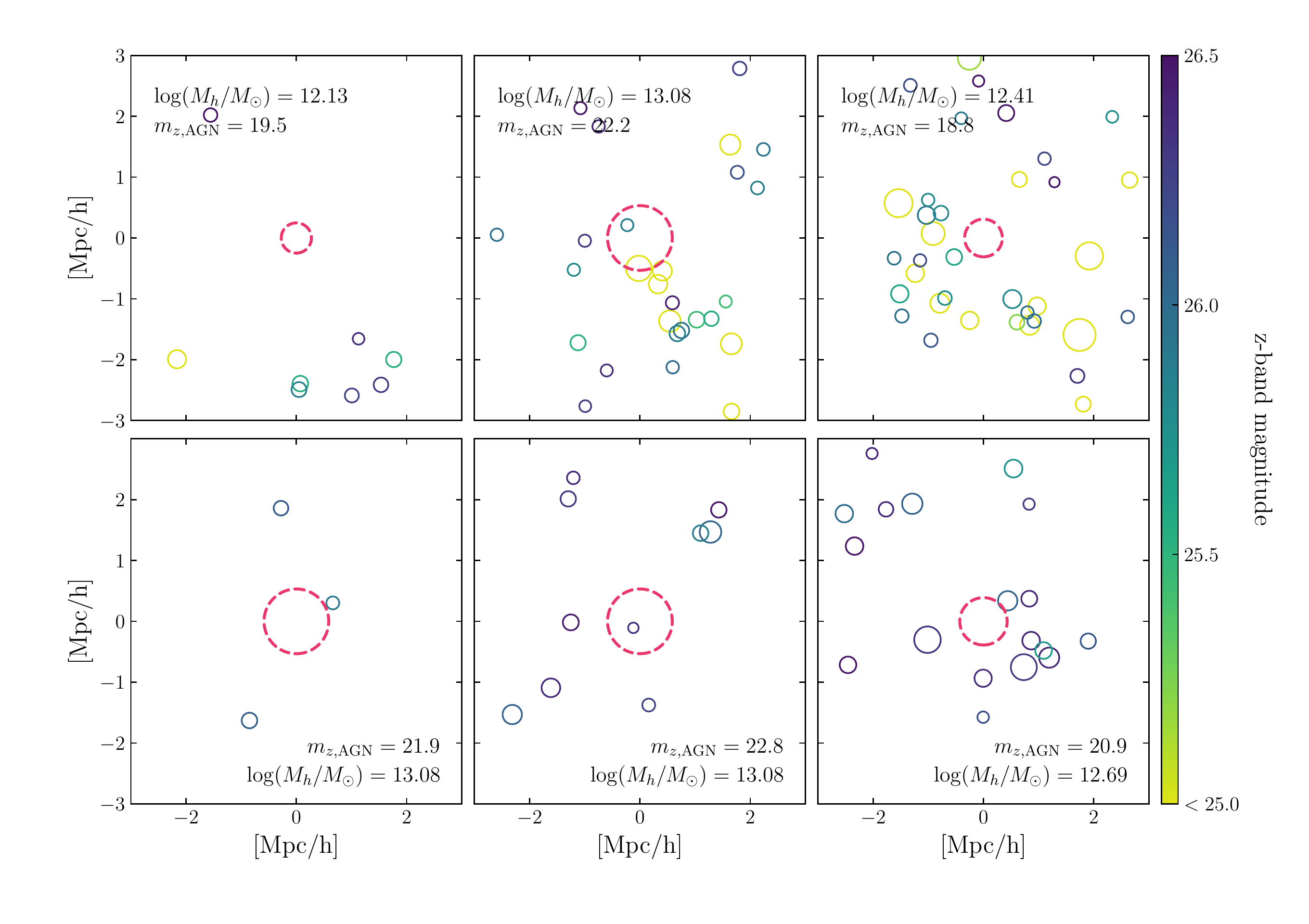}}
	\caption{\small Selected mock fields-of-view centred over a bright quasar showing the diverse quasar environments ordered by density from left to right. The top set of panels are outputs from the full hydrodynamical suite, BlueTides. The bottom set of panels are views from our empirical model with the parameters, $\Sigma_{G} = 0.3$, $\Sigma_{Q} = 0.5$ and $\varepsilon_{DC} = 1$. Note that the BlueTides set of results are not dust corrected, hence the number counts and magnitudes computed may be overestimated. Both magnitude and halo mass of the central quasar are annotated at the top left (lower right) of the top (lower) panels. The colours show the luminosities of galaxies (solid outline) with the flux limit set at $m_{z} < 26.5$. The central quasar (red, dashed outline) has its z-band magnitude explicitly stated. The size of the objects are exaggerated to represent changes in log halo mass, and is not an indication of the actual halo extent.}
	\label{fig:fig0}
\end{figure}

\begin{figure}[ht!]
	\centerline{\includegraphics[angle=-00, scale=0.65]{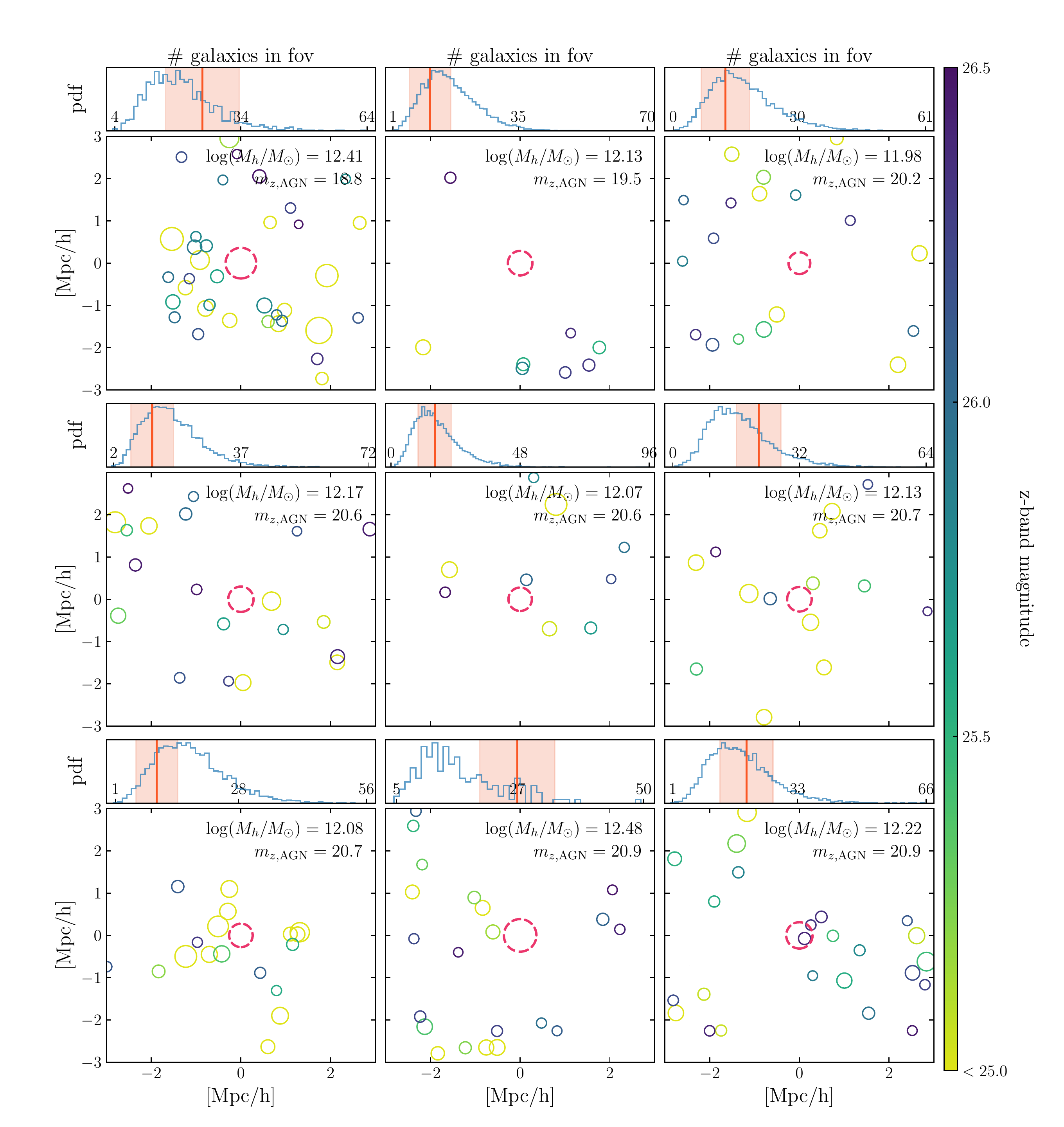}}
	\caption{\small Same as Fig.~\ref{fig:fig0}, but random fields-of-view from the BlueTides simulation centred around the $9$ most luminous quasars. The histogram represents the distribution of neighboring galaxy number counts from $1 \times 10^{2}$ randomly oriented pencil beams for a set of similar mass halos ($\Delta \log(M_{h}) = 0.1$) with lower luminosity AGNs (fainter than $3$ mag relative to the targeted quasar) in the simulation. The shaded region in the histogram corresponds to the average and the associated 1$\sigma$ uncertainty in neighboring galaxy number counts for our targeted most luminous quasar. We note that the entire set of luminous quasars appear to be in similar density environments to their similar mass, fainter AGN counterparts.}
	\label{fig:fig05}
\end{figure}

\begin{figure}[h!]
	\centerline{\includegraphics[angle=-00, scale=0.60]{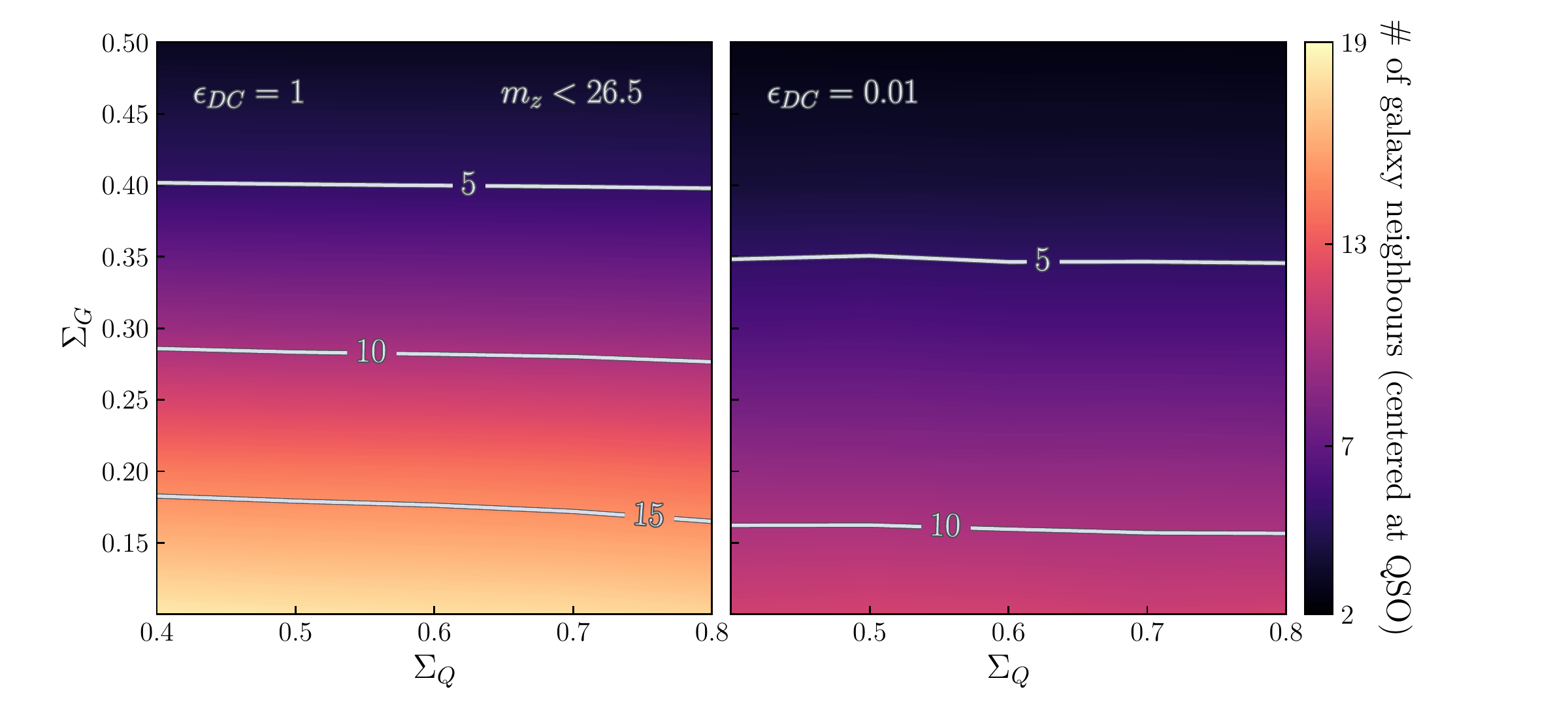}}
	\caption{\small The average number of visible galaxies ($m_{z} < 26.5$) in a quasar field as a function of scatter parameters, $\Sigma_{G}$ and $\Sigma_{Q}$ assuming duty cycle values, $\varepsilon_{DC} = 1$ (left) and $\varepsilon_{DC} = 0.01$ (right). As with \citet{Kim_2009}, the galaxy neighbour count excludes the central galaxy that is contains our bright quasar. }
	\label{fig:dcboth}
\end{figure}

\begin{figure}[h!]
	\centerline{\includegraphics[angle=-00, scale=0.64]{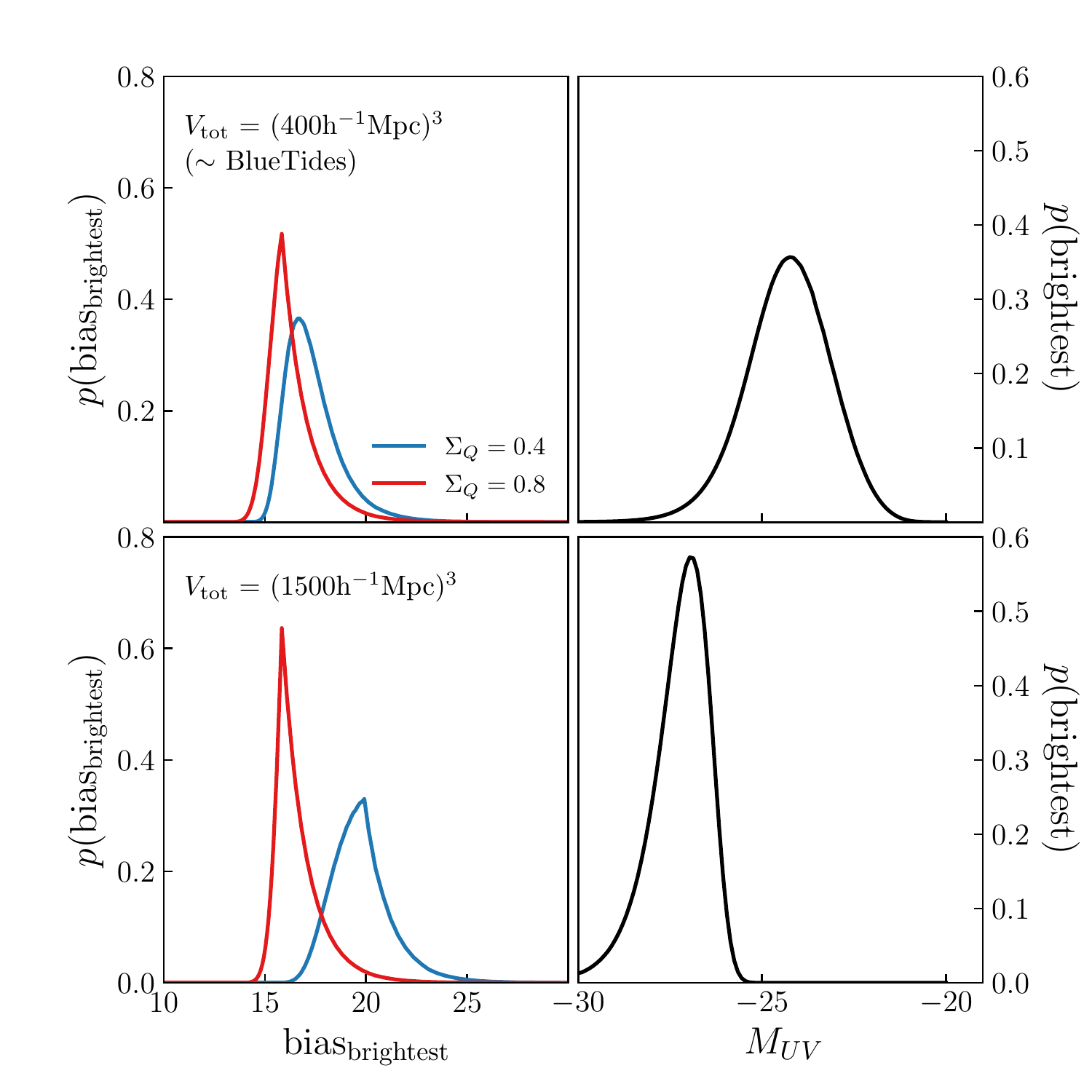}}
	\caption{\small Modeling the probability distribution of the bias for the brightest object, $p(\mathrm{bias}_{\mathrm{brightest}}$) inside a simulation volume, $V_{\mathrm{tot}}$ of ($400\mathrm{h}^{-1}$Mpc)$^{3}$, corresponding to the catalogue used in this work, BlueTides (upper left) and hypothetical N-body simulation with a larger volume ($1500\mathrm{h}^{-1}$Mpc)$^{3}$ (lower left). For each of these volumes, we look at the two limiting cases of $\Sigma_{Q} = 0.4$ (solid blue) and $\Sigma_{Q} = 0.8$ (solid red). The set of panels on the right show the distribution of magnitudes for the brightest quasar inside their respective volumes (BlueTides, upper right; larger simulation, lower right) by solving Equation~\ref{eqn:mag}.}
	\label{fig:bias}
\end{figure}

\begin{figure}[h!]
	\centerline{\includegraphics[angle=-00, scale=0.60]{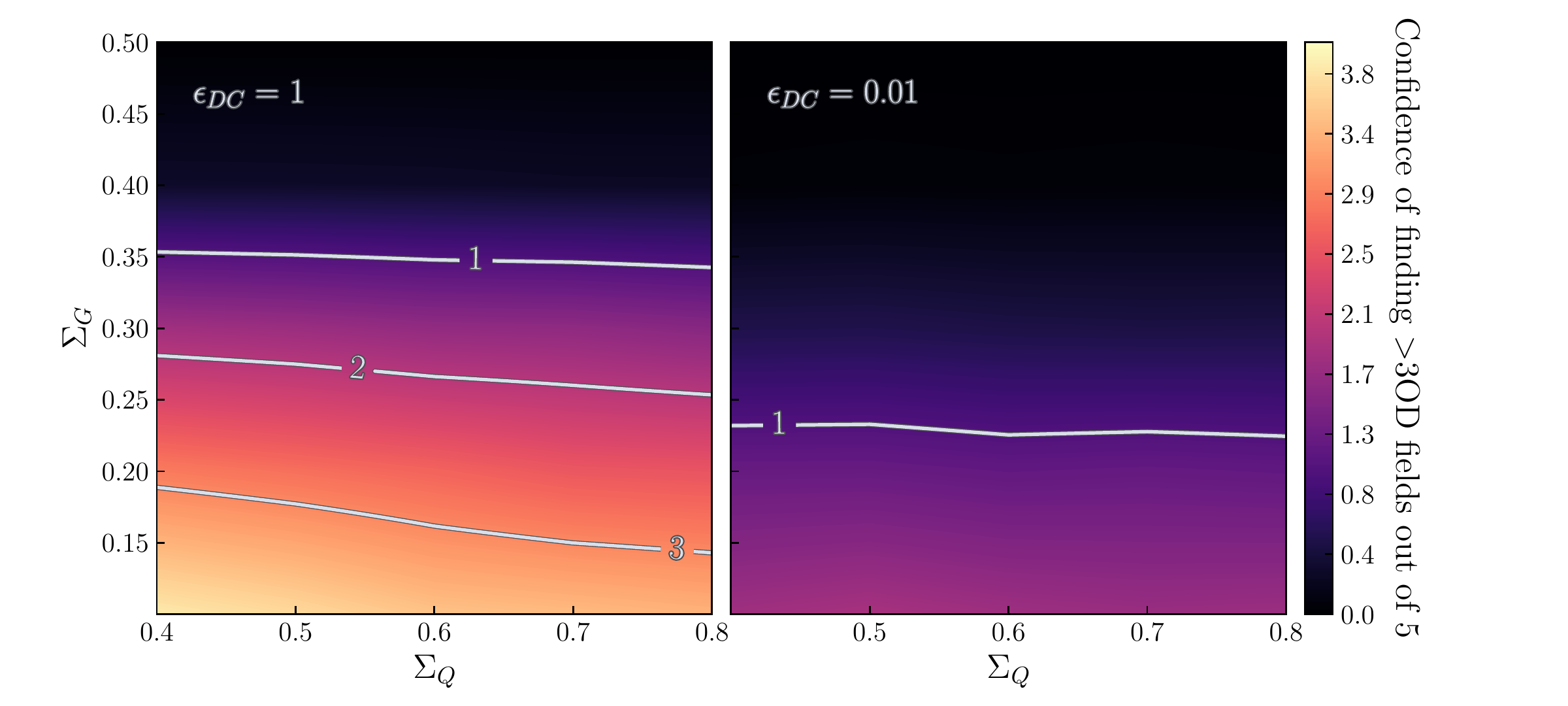}}
	\caption{\small The confidence in standard deviation of finding at least $3$ overdense (OD) quasar fields in a set of $5$ as a function of scatter parameters, $\Sigma_{G}$ and $\Sigma_{Q}$. An overdense field is a field with greater than $>4$ visible galaxies ($m_{z} < 26.5$), where an underdense quasar field would have $<=4$ galaxies. Different duty cycles $\varepsilon_{DC} = 1$ (left) and $\varepsilon_{DC} = 0.01$ (right). The $2\sigma$ contour represents a $0.05$ probability of $5$ drawing quasar fields matching the set from \citet{Kim_2009}.}
	\label{fig:dcboth2ud5}
\end{figure}

We summarize our key results below:

\begin{itemize}

\item We show that for $\varepsilon_{DC} = 1$, our model converges to the interpretation that rare luminous quasars reside in overdense halo environments, with hosts having halo bias of order $b \sim O(10^{1})$ (Fig.~\ref{fig:bias}). Furthermore, the high halo bias value persists irrespective of how high we set $\Sigma_{Q}$. However, we find that even with similar halo biases, the galaxy number counts between field-to-field can still vary significantly (Fig.~\ref{fig:fig0}). Fundamentally, this suggests the diversity in clustering of luminous galaxies around a quasar host is driven by galaxy evolution processes in neighbouring halos instead of the large scale structure phenomenology of the underlying density field. This is consistent with the findings of \citet{Costa_2014} also alluding that galaxy feedback processes play a major role.

\item In our modeling, the extent of visible galaxy clustering is sensitive to $\Sigma_{G}$ and $\varepsilon_{DC}$. The variation in galaxy luminosities set by $\Sigma_{G}$ directly increases the variance of visible galaxy counts in a quasar field. $\Sigma_{G}$ also implicitly sets a scale for radio-mode AGN feedback where massive neighbouring halos are increasingly likely to have self-regulated star formation, thus facilitating the odds for a underdense quasar field. Similarly, the impact of changing $\varepsilon_{DC}$ would alter the distribution of halo masses that hold quasars. Decreasing $\varepsilon_{DC}$ reduces the probability for a SMBH to be accreting in any instance would skew the most luminous quasar towards the less-dense common halos as $\varepsilon_{DC}$ is mass-independent. 

\item Setting $\varepsilon_{DC} = 0.01$ potentially relieves tension in the weak clustering around quasar fields, however this can be difficult to justify on a theoretical basis of achieving $10^{9} M_{\odot}$ SMBHs by $z \sim 7$. In addition, $\varepsilon_{DC}$ can be difficult to constrain as it is sensitive to obscuration effects potentially underestimating the value \citep{Chen_2018, Trebitsch_2019}. $\Sigma_{G}$ adds to this by suggesting an unaccounted population of feedback-affected galaxies beyond the depth of the observation window.

\item We also see that $\Sigma_{Q}$ has a small/no impact on the clustering of bright objects. We find the reason is related to the simulation volume used (Fig.~\ref{fig:bias}). For low $\Sigma_{Q}$, the BlueTides N-body simulation did not have the necessary volume to probe enough halos past $M_{h} > M_{h}^{c}$. However, we expect the overall impact from this effect to be fairly small even with sufficient volume as there is significant overlap in the bias distribution across the range of $\Sigma_{Q}$.

\item We find that clustering eventually becomes independent on quasar luminosity (Fig.~\ref{fig:bias}). This is a natural consequence of assuming some feedback scale in $L_{Q,c}(M_{h})$. In our model, the median quasar luminosity does not change past $M_{h} > M_{h}^{c}$. As the number density of massive halos drop exponentially, we expect the most luminous quasars to be tend towards having a bias value, $b(M_{h}^{c})$. 

\item In Figs.~\ref{fig:dcboth} and \ref{fig:bias}, we show that our results are in agreement with existing simulations finding quasar environments to be generally overdense with a high degree of variance in galaxy counts \citep{RomanoDiaz_2011, Costa_2014, Habouzit_2019}. However, there is marginal difficulty for these simulations to reproduce underdense fields of \citet{Kim_2009}. Justifications include invoking a low Lyman-break galaxy duty cycle \citep{RomanoDiaz_2011}, strong galactic winds \citep{Costa_2014} or tentatively, radiative feedback from neighbouring quasars \citet{Habouzit_2019}.

\item Our model requires $\Sigma_{G} \sim 0.3$ to remain consistent with the observations of \citet{Kim_2009} (Fig.~\ref{fig:dcboth2ud5}). Additionally, current state-of-the-art models on high redshift galaxy evolution (e.g. \citealt{Mutch_2016, Tacchella2018, Behroozi2019, Vogelsberger_2020} and BlueTides) all tend close to this value of $\Sigma_{G}$, suggesting reported underdense fields are not necessarily unlikely events.

\item One limitation to note is the independent placement of galaxies and quasars in halos neglects any large scale (2-halo) environmental effects. Currently, there is limited evidence that suggest UV quasar radiation can be intense enough to suppress star formation in nearby central galaxies, albeit in relatively small halos $M_{h} < 10^{9} M_{\odot}$ \citep{Kashikawa_2007}. \citet{Habouzit_2019} suggests that the quasar mode radiation from the AGN is capable of ionizing neutral hydrogen gas up to $\sim 10$cMpc. However, they contend that more detailed radiative transfer simulations are required to measure the significance of this type of feedback on galaxy growth. In the context of our model any additional feedback would further serve as another source to relieve the tension between the observations and our modeling.

\end{itemize}

\acknowledgements
This research was conducted by the Australian Research Council Centre of Excellence for All Sky Astrophysics in 3 Dimensions (ASTRO 3D), through project number CE170100013. K.R and M.M are additionally supported through the Research Training Program Scholarship from the Australian Government. K.R acknowledges support through the Postgraduate Writing-up Award granted by the David Bay Fund. MM acknowledges support from the National Research Council of Canada’s Plaskett Fellowship.

%=====

\bibliographystyle{aasjournal}
\bibliography{qsopaper}

\appendix
\section{Quasar Duty Cycle in BlueTides} \label{apdx:a}

\begin{figure}[h!]
	\centerline{\includegraphics[angle=-00, scale=0.70]{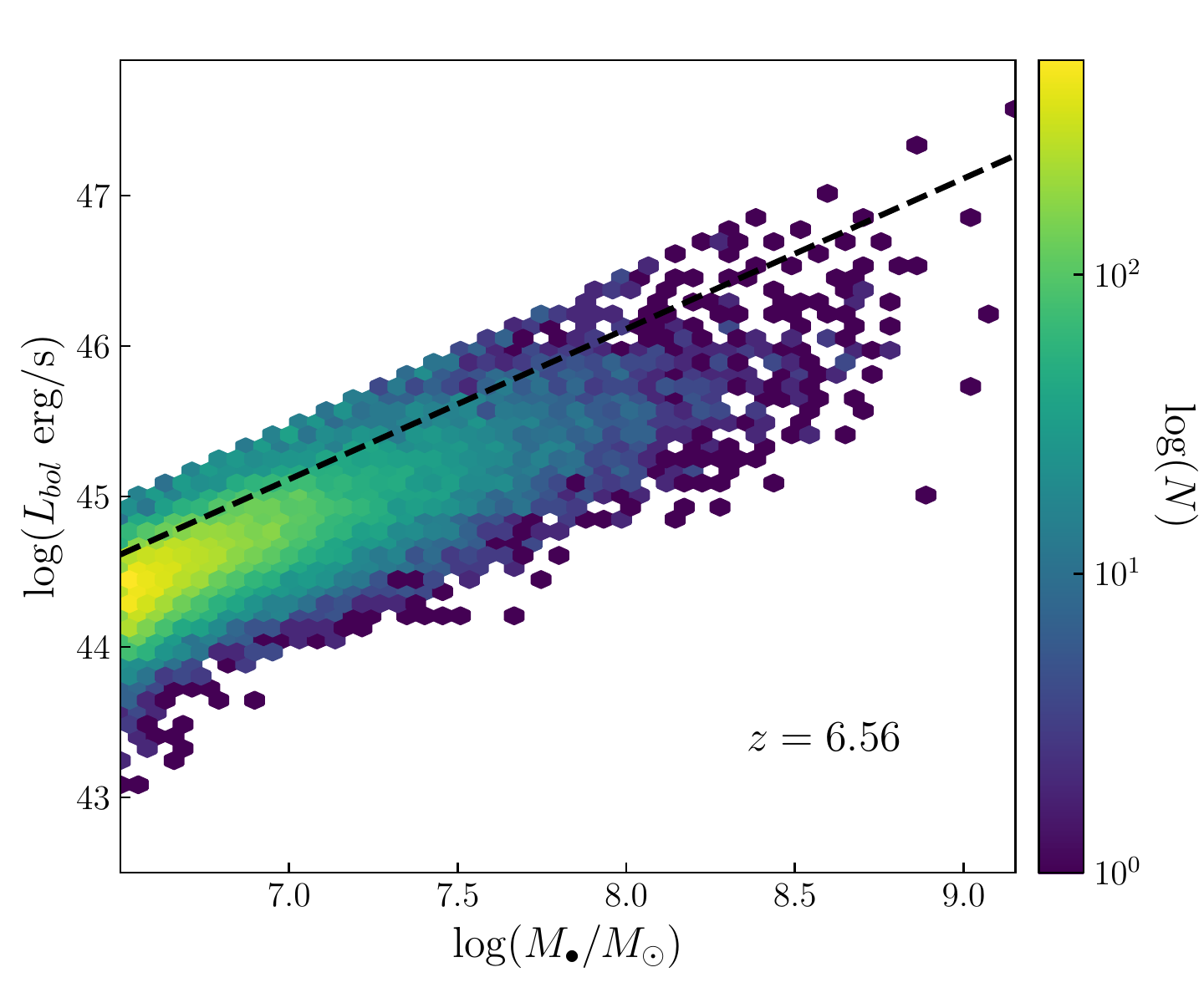}}
	\caption{\small AGN luminosity, $L_{bol}$ as a function of SMBH mass, $M_{\bullet}$ in BlueTides at $z=6.56$. The dotted black line is the Eddington luminosity. We note that all SMBHs in BlueTides are undergoing accretion.}
	\label{fig:apdxA}
\end{figure}

Black holes in BlueTides uses the sub-grid model developed for its predecessors: MassiveBlack I \& II \citep{Springel_2005, Di_Matteo_2005}, with additional modifications consistent with Illustris \citep{DeGraf_2016}. Black holes are initially seeded with a mass of $M_{\bullet} = 5\times10^{5}$h$^{-1}M_{\odot}$ in halos more massive than $M_{h} = 5 \times 10^{10}$h$^{-1}M_{\odot}$. In BlueTides, black holes accrete gas via Bondi-Hoyle accretion, 

\begin{equation*}
\dot{M_{\bullet}} = \dfrac{4\pi \alpha G^{2}M_{\bullet}^{2}}{\rho(c_{s}^{2} + v^{2})^{3/2}},
\label{eqn:bondihoyle}
\end{equation*}

where $\alpha$ is a dimensionless parameter, $G$ is the gravitational constant, $\rho$ is the local density of gas, $c_{s}$ is the local sound speed and $v$ is the relative velocity between the black hole and the nearby gas. BlueTides also allows for super-Eddington accretion, with a maximum permitted rate of $2$ times the Eddington limit,

\begin{equation*}
\dot{M}_{Edd} = \dfrac{4\pi G M_{\bullet} m_{p}}{\eta \sigma_{T} c},
\label{eqn:edd}
\end{equation*}

where $m_{p}$ is the proton mass, $\sigma_{T}$ is the Thomson cross-section of an electron and $\eta = 0.1$ is the radiative efficiency. The conversion from accretion rate to bolometric luminosity, $L_{bol}$ is given as,

\begin{equation*}
L_{bol} = \eta \dot{M_{\bullet}}c^{2}, 
\label{eqn:radeff}
\end{equation*}

where $c$ is the speed of light. In Fig.~\ref{fig:apdxA} we show the distribution of SMBH masses and their AGN luminosities in BlueTides at $z=6.56$. We select for established black holes using the criteria, $M_{\bullet} > 10^{6.5} M_{\odot}$ to avoid any possible uncertainty originating from the BlueTides seeding prescription. For the quasar duty cycle, we follow the definition provided in \citet{DeGraf_2016} and compute the fraction of quasars that exceed some selected luminosity threshold. The calculated quasar duty cycle is, $\varepsilon_{DC} \sim 1$ for $L_{bol} > 10^{44}$erg$/$s, consistent with the projection for the $z > 6$ duty cycle of \citet{DeGraf_2016} using Illustris. Additionally, $86\% (53\%)$ of black holes in BlueTides have AGN luminosities brighter than $L_{bol} > 10^{44.3 (44.5)}$erg$/$s, corresponding to $50\% (75\%)$ of the Eddington limit for a $M_{\bullet} = 10^{6.5} M_{\odot}$ black hole.

\section{Galaxy Luminosity Scatter in BlueTides} \label{apdx:b}

\begin{figure}[h!]
	\centerline{\includegraphics[angle=-00, scale=0.70]{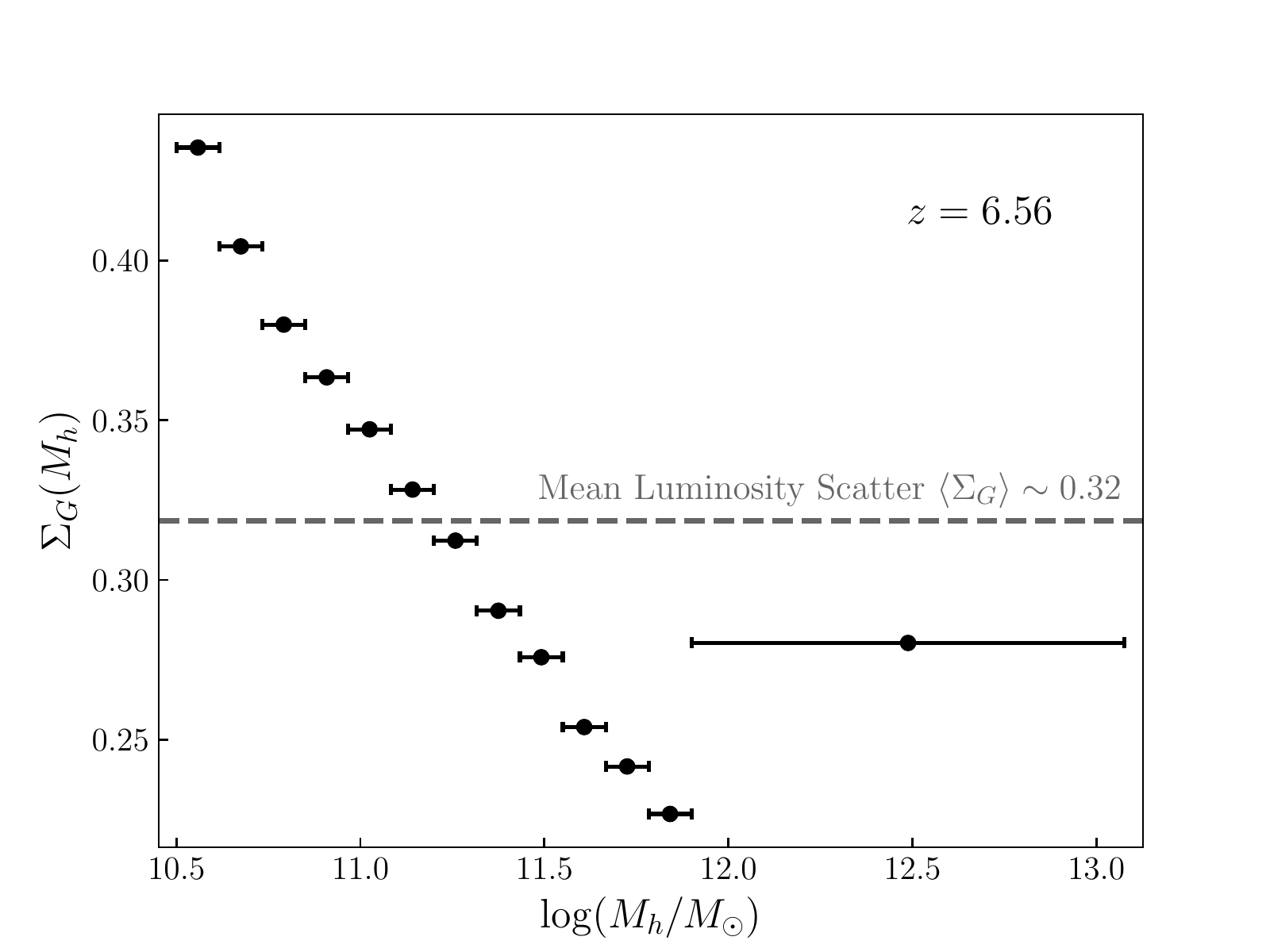}}
	\caption{\small Dispersion in galaxy luminosities, $\Sigma_{G}$ as a function of halo mass, $M_{h}$ in BlueTides at $z=6.56$. Halo masses are binned in intervals of $\Delta(\log M_{h}) = 0.12$, with the exception of the final bin centred at $M_{h} \sim 10^{12.5}M_{\odot}$ with a bin size of $\Delta(\log M_{h}) \sim 1$. The mean luminosity scatter across all bins is $\langle \Sigma_{G} \rangle \sim 0.32$. }
	\label{fig:apdxB}
\end{figure}

We compute the scatter in galaxy luminosities, $\Sigma_{G}$ in various halo mass bins of size $\Delta(\log M_{h}) =0.12$ in BlueTides at $z=6.56$. The final mass bin contains all halos $M_{h} \gtrsim 10^{12}M_{\odot}$ and is selected in a way such that the number of objects is comparable in size to the mass bin before it. The galaxies in BlueTides are not corrected for dust, hence the scatter values are underestimated as the inclusion of dust will add an additional degree of variability. In BlueTides, we find the mean luminosity scatter across all bins is $\langle \Sigma_{G} \rangle \sim 0.32$, which is comparable in magnitude to the hydrodynamical simulation IllustrisTNG at the same redshift, with $\langle\Sigma_{G} \rangle  > 0.3$ (estimated from scatter in UV luminosity versus stellar mass relation; \citealt{Vogelsberger_2020}). The trend of decreasing $\Sigma_{G}$ as $M_{h}$ increases is consistent with both IllustrisTNG, and with the semi-analytical model, Meraxes for $z\sim 8$ galaxies \citep{Ren_2018}.

\end{document}